\documentclass[twocolumn,secnumarabic,amssymb,floatfix,aps,prl]{revtex4}

\usepackage{graphicx}
\usepackage{dcolumn}
\usepackage{bm,amsmath,amsbsy}     
\usepackage{braket}
\usepackage{appendix}
\usepackage{xcolor}
\usepackage{array}
\usepackage{booktabs}
\usepackage{soul}
\setstcolor{red}
\usepackage{amsthm}
\usepackage{amsfonts}
\usepackage{float}
\usepackage{pstricks}           
\usepackage{mathrsfs}
\usepackage{hyperref}           
\usepackage{hypernat}           
\usepackage[latin1]{inputenc}   
\usepackage{subfigure}
\usepackage{slashed}    
\usepackage{color}
\usepackage{soul}

\begin{document}
\title{Quantum pathways interference in laser-induced electron diffraction revealed by a novel semiclassical method} 

\author{Phi-Hung Tran$^1$}
\author{Van-Hung Hoang$^2$}
\author{Anh-Thu Le$^1$}
\affiliation{$^1$Department of Physics, University of Connecticut, 196A Auditorium Road, Unit 3046, Storrs, CT 06269}
\affiliation{$^{2}$J. R. Macdonald Laboratory, Department of Physics, Kansas State University, Manhattan, Kansas 66506, USA}
\date{\today}

\begin{abstract}
We develop a novel method for strong-laser-field physics based on the combination of the semiclassical Herman-Kluk propagator and the strong-field approximation and demonstrate its high accuracy on the calculations of photoelectron momentum distribution (PMD) for atoms and molecules in intense lasers. For rescattered electrons, we show that for a given time that electron tunnels to the continuum, there are typically multiple trajectories that lead to the same final momentum in the high-energy region. These trajectories start with slightly different initial transverse momenta and carry different phases giving rise to the interference structures in the PMD, which can also be associated with the laser-free electron-ion differential cross section. This is in contrast to the well-known long and short trajectories, which result in different interference patterns. Our results can be used to extend current capabilities of the laser-induced electron diffraction and other ultrafast imaging and strong-field spectroscopic techniques.

\end{abstract}
\maketitle

Semiclassical approach has a long and successful history of providing physical insight and quantitative approximations to various problems in quantum physics \cite{Heller:jcp75,Miller:jcp70,Child:book2018}. In strong-field physics, early applications of semiclassical approach were very successful in providing an intuitive picture of the three-step model for the description of important recollision phenomena \cite{Corkum:prl93,Schafer:prl93,Lewenstein:pra94,Lewenstein:pra95}. Within this model, the active electron first tunnels through the combined potential of the atom and the laser electric field. It then propagates in the continuum, and, as the laser electric field changes its direction, it has a chance to revisit the parent ion and recombine with the core with an emission of a high-energy photon, or elastically scatter from it. There are strong needs for accurate description and numerical simulations for these processes as they are at the heart of modern ultrafast strong-field imaging techniques such as high-harmonic spectroscopy \cite{Smirnova:nature09,Worner:nature10,Worner:science11,Worner:science15} and laser-induced electron diffraction (LIED) \cite{Blaga:nature12,Wolter:science16,Pullen:NatCom15}. Nevertheless, numerical calculations based on this semiclassical method, so-called the strong-field approximation (SFA), only show semi-quantitative agreements with exact numerical solutions of the time-dependent Schr\"odinger equation (TDSE) \cite{Krausz:rmp09,Lin:book2018,Milosevic:jpb06,Faria:PhysRep20}. To be specific, in this Letter we will focus on the photoelectron spectra.

The main reason for the inaccuracies of the SFA can be traced back to the fact that the interaction between electron in the continuum and atomic core is mostly neglected. Over the past two decades, various improvements over the SFA has been proposed to account for the Coulomb effect in the continuum ~\cite{Popruzhenko:jpb14,Popruzhenko:prl08,Smirnova:pra08,Cerkic:pra09}. Another approach is to solve classical equations for electron motion in the combined fields and incorporate their effects in the action so that the quantum interference is taken into account ~\cite{Yan:prl10,Li:prl14,Shvetsov:pra16,Faria:PhysRep20}. Most improvements have been achieved so far are for the low-energy photoelectron momentum distribution (PMD). Despite this progress, in general, these methods show only fair agreements with the TDSE solutions for the rescattered electrons in the high-energy region. 

The difficulties in obtaining accurate high-energy PMD can be associated with the use of the Van Vleck -- Gutzwiller (VVG) propagator that most current applications of semiclassical methods rely on. Within the VVG, one must perform the root search to connect the final electron momenta with the initial conditions for all trajectories. Furthermore, the VVG also suffers from singularities at caustics. To simplify the problem, it has been a common practice to neglect the preexponential factor in the VVG propagator. Recently, Brennecke {\it et al.} \cite{Brennecke2020} showed that correct magnitude and phase, including the Morse (or Maslov) index, of the preexponential factor are needed to get accurate results. 
 
In this Letter, we propose a method that combines the Herman-Kluk (HK) semiclassical propagator \cite{Herman1984,Kluk1986} with strong-field formalism, to overcome the difficulties associated with the VVG propagator and show that excellent agreements with the TDSE can be achieved. Furthermore, we show that for any given final momentum in the rescattering region, there are multiple trajectories in general, even within a given laser cycle. These trajectories differ by their initial transverse momenta, which leads to different effective impact parameters upon recollisions. The interference among these trajectories results in specific structures in the PMD which can serve as ``fingerprints" of the target. As shown earlier in the quantitative rescattering theory (QRS) \cite{Morishita:prl08,Le:pra09,Lin:jpb10,Lin:book2018}, the signals in the high-energy PMD can be associated with the laser-free electron-ion elastic differential cross section (DCS). Our results therefore provide a clear physical interpretation for the factorization of the recollision process, established in the QRS. We remark that our finding is in a strong contrast to the common belief that there are only two, so-called long and short trajectories. 

Within our method, referred to herein as the strong-field Herman-Kluk (SFHK) approximation, the amplitude for the transition from the initial state $\left|\Phi_0\right\rangle$ to the final state with momentum ${\bf{p}}_f$ is first written using the HK propagator as (see the Appendix for more details)
\begin{equation}
\begin{aligned}
{\Psi}\left( {{\bf{p}}_f,t} \right) =&\frac{i}{(2\pi)^3} \int_{{t'}}^{{t}} {dt'} \int\!\!\!\int {d{\bf{p}}\,d{\bf{q}}} \,{C_{{\bf{pq}}t}}{e^{i{S}}} \ \   \\
&\times \left\langle {{{\bf{p}}_f}}
 \mathrel{\left | {\vphantom {{{\bf{p}}_f} {{{\bf{p}}_t},{{\bf{q}}_t}}}}
 \right. \kern-\nulldelimiterspace}
 {{{{\bf{p}}_t},{{\bf{q}}_t}}} \right\rangle \left\langle {{\bf{p}},{\bf{q}}\left| {{\bf{r}}\cdot{\bf{E}}\left( {{t'}} \right)} \right|{\Phi _0}} \right\rangle.
\end{aligned}
\label{SFHKtWF}
\end{equation}
Here, $({\bf{p}}_t,{\bf{q}}_t)$ represents the phase space location at time $t$ of a classical trajectory that starts at $(\bf{p},\bf{q})$ at time $t'$, $S$ is the action along this trajectory, $\left|{\bf{p},\bf{q}}\right\rangle$ is the coherent state with average momentum $\bf{p}$ and position $\bf{q}$. 

\begin{figure} [htb]
	\centering
        \includegraphics[width=0.9\linewidth]{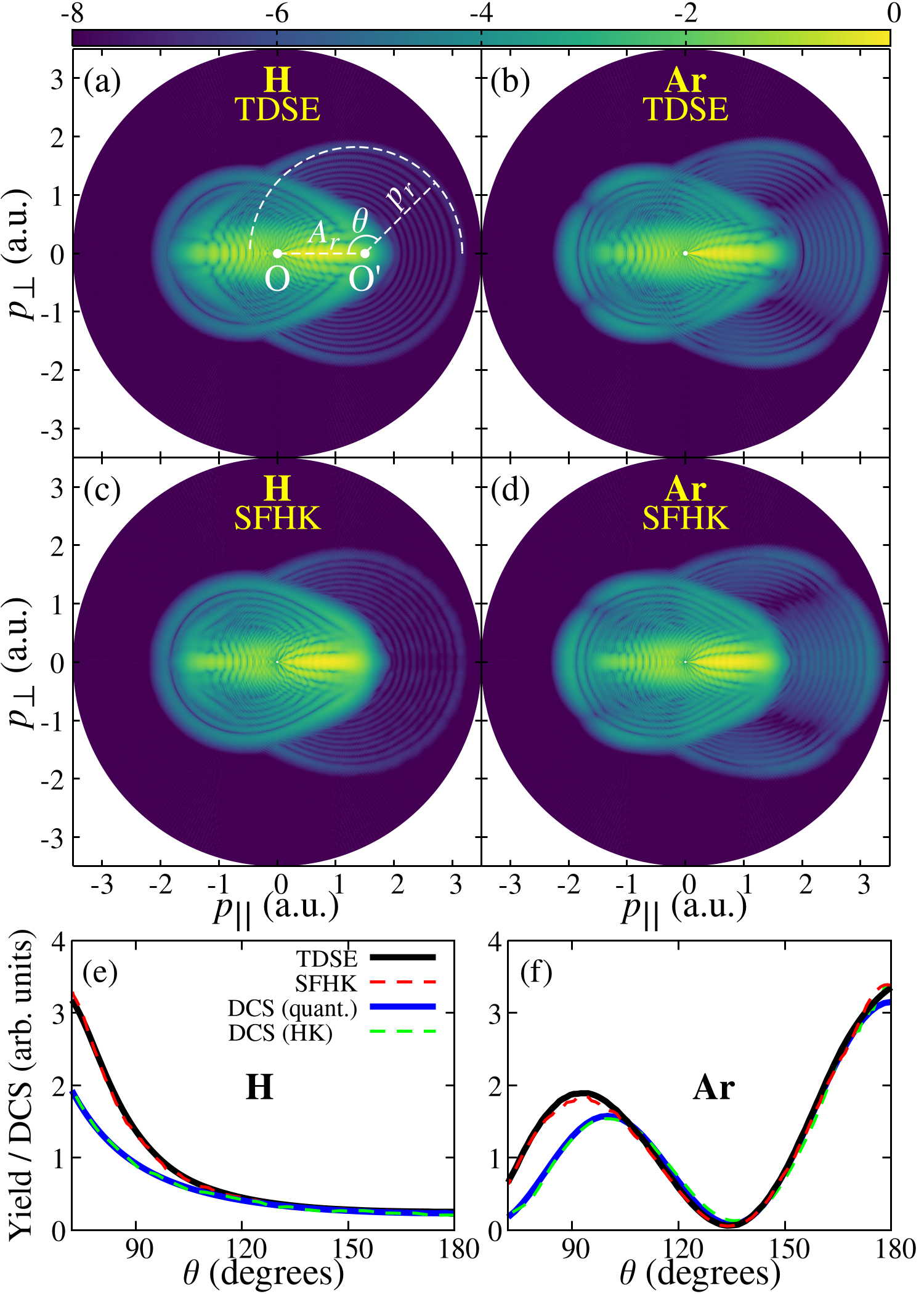}
	\caption{(a) and (c): 2D PMD for H obtained from the TDSE and our SFHK method, respectively. A linearly polarized, three-cycle, 1200-nm wavelength laser pulse with an intensity of $10^{14}$ W/cm$^2$ was used. (b) and (d): same as (a) and (c), but for Ar. (e) and (f): PMD for H and Ar, respectively, along the white dashed semi-circle indicated in (a). Also shown are the electron-ion laser-free elastic scattering DCS for an incident momentum $k = 1.8$ a.u. calculated with a fully quantum and HK methods. The radius and center of the semi-circle are defined by $p_r = 1.8$ a.u. and $A_r = 1.35$ a.u., which are identical to the electron momentum and A-vector at the closest approach, respectively (see Fig.~\ref{Fig2}).}
	\label{Fig1}
\end{figure}

The physical meaning of Eq.~(\ref{SFHKtWF}) is clear. The last factor on the right hand side gives the part of the initial wavefunction tunneled to the continuum at time $t'$, projected onto the coherent state centered at $({\bf{p}},{\bf{q}})$. Each coherent state is then propagated along a classical trajectory, multiplied by the factor $Ce^{iS}$. The final wavefunction is obtained by summing over all the contributions from those coherent states at the final time which can be carried out by Monte-Carlo integration with importance sampling. Although Eq.~(\ref{SFHKtWF}) was proposed earlier ~\cite{Spanner:prl03,Walser:jpb03}, it has not been used in practice. A possible reason could be that it involves a seven-fold integral. Instead of using Eq.~(\ref{SFHKtWF}) directly, we propose to approximate the tunneling at time $t'$ by the SFA in the saddle-point approximation. The ``tunnel exit'' is given by ${{\bf{q}}_{t'}} = {\mathop{\rm Re}\nolimits} \left[ {\int_{{t_s}}^{{t'}} {{\bf{A}}\left( t \right)dt} }\right]$, with $t_s=t'+it_t$ 
and $t_t$ being the tunneling time \cite{Brennecke2020,Yan:prl10,Ivanov:jmo05}. For each $t'$ we associate a trajectory starting at $({\bf{p},\bf{q}})$ with a coherent state $\left|{\bf{p},\bf{q}}\right\rangle$ weighted by the SFA ionization probability amplitude. The SFA has been used earlier in combination with the VVG propagator ~\cite{Brennecke2020,Faria:PhysRep20,Yan:prl10}, but without any association with coherent states. Within the SFHK, the transition amplitude is written as
\begin{equation}
{M_{{\bf{p}}_f}} \propto \sum_{t_s}\sum_{({\bf{p},\bf{q}})}  \left\langle {\bf{p}}_f
 \mathrel{\left | {\vphantom {{\Psi _C^{{\bf{p'}}}} {{{\bf{p}}_t},{{\bf{q}}_t}}}}
 \right. \kern-\nulldelimiterspace}
 {{{{\bf{p}}_t},{{\bf{q}}_t}}} \right\rangle {C_{{\bf{pq}}t}}{e^{i\left( {S_ \downarrow ^0 + {S_ \to }} \right)}}.
\label{Mp_main}
\end{equation}
Here, the action is split into two terms, $S\left( {{t_s}} \right) = S_ \downarrow ^0 + {S_ \to }$, associating with tunneling and propagating, respectively \cite{Brennecke2020,Lai2015}.

As compared with the direct use of Eq.~(\ref{SFHKtWF}), our method reduces the integral effectively to a 3D integral, as $({\bf{p},\bf{q}})$ satisfy the saddle-point equations. As compared with the methods that rely on the VVG propagator, our method inherits several advantages from the HK propagator. First, as an initial value representation \cite{Miller:jcp70,Miller:jpcA01} instead of a boundary value problem, the root search is completely avoided. The root search becomes increasingly difficult when the system experiences chaotic behaviors, as noticed earlier \cite{Sand:prl99,Sand:pra00}. Second, the HK propagator was shown to be a uniform approximation \cite{Kay:ChemPhys06}, whereas within the VVG, one has to deal with the singularities of the preexponent factor (i.e., the caustics). 


As an illustration, we compare in Fig.~\ref{Fig1} the PMDs obtained by the SFHK with the TDSE results for H and Ar. The agreement between the two methods is astounding in the whole range of spectra. In the following we will focus on the high-energy part seen as concentric rings in the PMD that is originated from rescattered electrons. The origin of these rings, in particular, the minima seen in the rings for the Ar case, has been understood based on the QRS theory \cite{Morishita:prl08,Lin:jpb10,Lin:book2018}. Very good agreements with the TDSE were found for other targets with different laser parameters -- see the Supplemental Material Fig.~S1. This level of accuracy for target-specific structures in the high-energy PMD has never been reported earlier with semiclassical methods.

\begin{figure} [thb]
	\centering
        \includegraphics[width=1.0\linewidth]{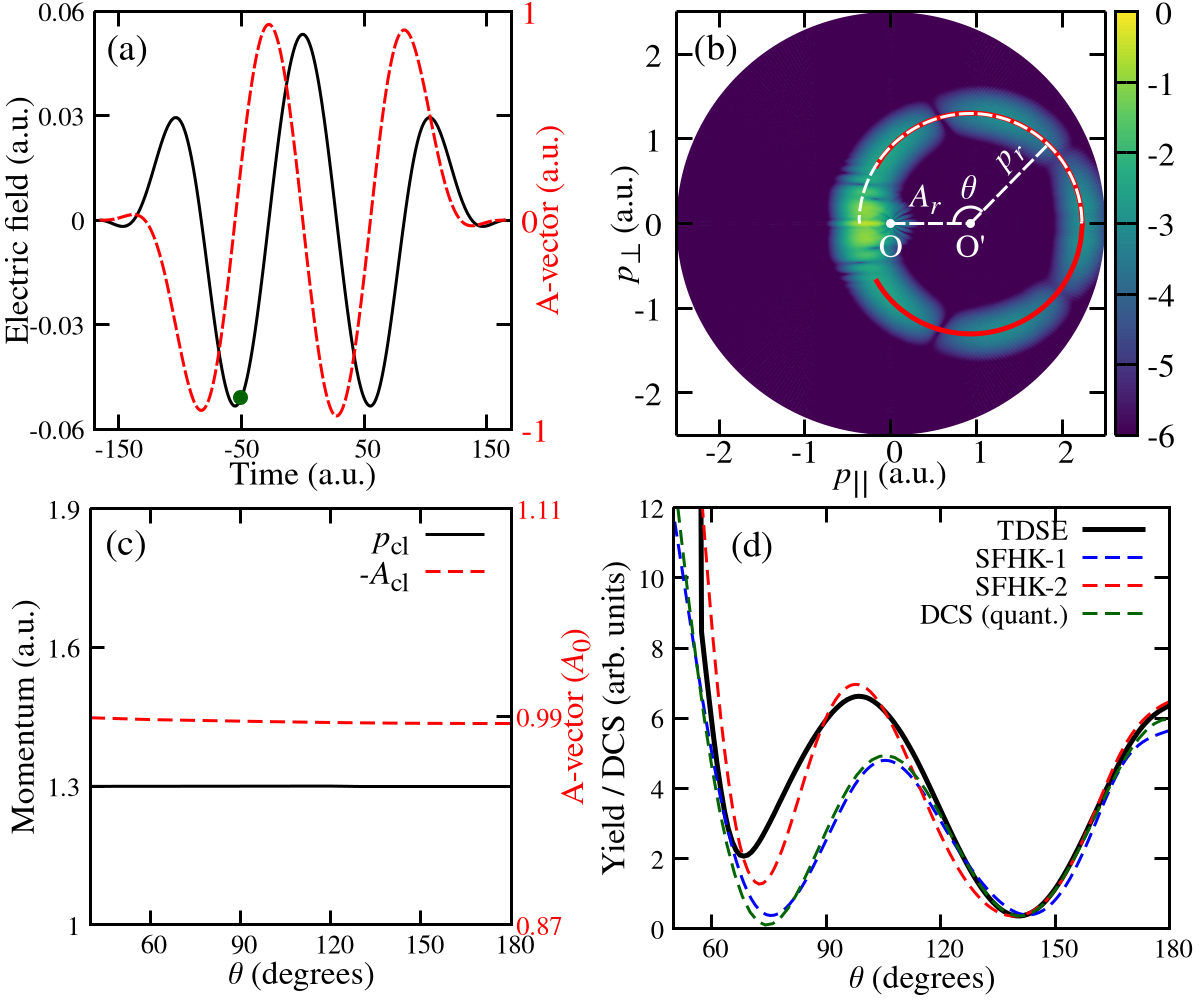}
	\caption{(a): Flattop laser pulse used in the subsequent analyses. See the main text for details. (b): The PMD obtained with the SFHK for Ar, in which the electron is allowed to tunnel to the continuum only in a narrow time interval with the phase $\varphi=\omega t' \in [7.69^{\circ}:7.71^{\circ}]$, marked by the green dot in (a) -- to be called SFHK-1. Here, $\varphi=0$ is defined at the sub-cycle peak near $t=-50$ a.u. (c): A-vector and electron momentum at the closest approach vs scattering angle in SFHK-1. (d): Comparison of the PMD along the white dashed semi-circle indicated in (b) by different methods. In SFHK-2 the tunneling window is $\varphi \in [5^{\circ}:8^{\circ}]$. The DCS for $k=1.3$ a.u. is also shown.}
	\label{Fig2}
\end{figure}

To have a quantitative comparison, we shown in Fig.~\ref{Fig1}(e)-(f) photoelectron yields along the ring with the highest energy [the white dashed semi-circle in Fig.~\ref{Fig1}(a)]. The center and the radius of the ring were determined using the QRS (see also discussion below). The agreements between the TDSE and our method are excellent for both targets. According to the QRS, the yield along this ring is factorized into a product of a returning electron wave-packet and the DCS with the incident momentum equal to the electron momentum at the time of recollision. We therefore also show in Fig.~\ref{Fig1}(e)-(f) the DCS obtained with the HK propagator together with the DCS from a fully quantum calculation. The agreements between the two methods are excellent for both targets. The good agreement between the photoelectron yields and the DCS for $\theta \gtrapprox 100^{\circ}$ indicates the adequacy of the QRS factorization in this range \cite{Morishita:prl08,Lin:jpb10}. We emphasize that the agreements between the SFHK and the TDSE remain excellent in the whole range for all the targets (see also Fig.~S2 of the Supplemental Material).

Does the discrepancy between the yields and the DCS seen in Fig.~\ref{Fig1} (e)-(f) imply the failure of the factorization below $\theta \approx 100^{\circ}$? To simplify our analysis in the following, we now analyze the case of Ar in a laser pulse of one-cycle flattop envelope with one-cycle turn-on (turn-off) ramps, as shown in Fig.~\ref{Fig2}(a). The laser is with the wavelength of 800~nm and intensity of $10^{14}$~W/cm$^2$. Fig.~\ref{Fig2}(b) shows the PMD obtained with the SFHK when the electron is allowed to tunnel to the continuum in a narrow time interval with  the phase $\varphi=\omega t' \in [7.69^{\circ}:7.71^{\circ}]$, marked as the green dot in Fig.~\ref{Fig2}(a). This was chosen to get the highest electron momentum after rescattering, in which the long and short trajectories merge together. Note that this phase is somewhat different from the simple classical estimate near $14^{\circ}$, when the core potential is neglected \cite{Chen:pra09}. The PMD obtained by this method (to be called SFHK-1) is spread around the classical final electron momenta [the red dots, which altogether show up as the red curve in the Fig.~\ref{Fig2}(b)].

\begin{figure}
	\centering
        \includegraphics[width=1.0\linewidth]{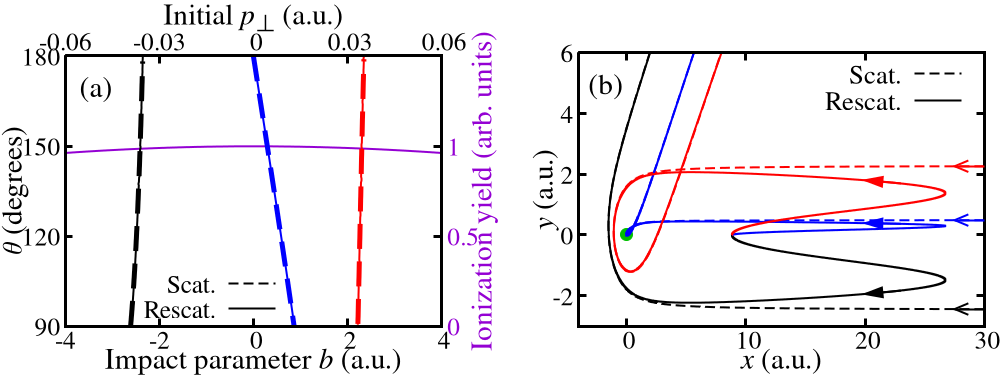}
	\caption{(a): Scattering angle vs impact parameter for e-Ar$^+$ collision at $k=1.3$ a.u. together with the rescattering case for three types of trajectories. See text for more details. Ionization probability vs initial $p_{\perp}$ is also shown. (b): Example of three trajectories with $k=1.3$ a.u. (for scattering) and $p_r=1.3$ a.u. (for rescattering) that leads to $\theta=130^{\circ}$. For rescattering case, the trajectories start near $x=10$ a.u. ($y=z=0$), while for the laser-free case, the electron is incident along negative $x$-axis from $x=200$ a.u., $y=b$, and $z=0$.}
	\label{Fig3}
\end{figure}

To understand the nature of the minimum structures seen in Fig.~\ref{Fig2}(b), we define the electron return time $t_{cl}$ for each trajectory as the time of the electron's closest approach to the ion core [see also Fig.~\ref{Fig3}(c)]. The vector potential at the return time, $A_\text{cl}$, is found to be nearly constant at about $99\%$ of its maximal value, see Fig.~\ref{Fig2}(c). This is consistent with an unique return time near about 3/4-cycle after ionization, i.e., the laser electric field nearly vanishes at $t_{cl}$. Electron momentum at the closest approach, defined as $p_\text{cl}=\sqrt{2E(t_\text{cl})}$, is also found to be nearly constant, see Fig.~\ref{Fig2}(c). Following the QRS, we identify $A_r \equiv  A_\text{cl}=0.93$~a.u. and $p_r \equiv  p_\text{cl}=1.3$~a.u. as the center and the radius of the white dashed ring shown in Fig.~\ref{Fig2}(b). This ring appears to be identical to the red curve, indicating the validity of our procedure for the determination of the location of the rescattered electrons in the 2D PMD, even when the full atomic potential is taken into account.

In Fig.~\ref{Fig2}(d) we compare the photoelectron yields from the SFHK-1 along the white dashed ring in Fig.~\ref{Fig2}(b) with the DCS for an incident momentum $k=p_r=1.3$~a.u. They agree very well down to $\theta\approx 60^{\circ}$. We found similar excellent agreements when other small tunneling windows were used, indicating the accuracy of the factorization in a broad range of $\theta$. However, as the tunneling window size increases, the factorization starts to deteriorate for small $\theta$, as illustrated in Fig.~\ref{Fig2}(d) for tunneling window $\varphi \in [5^{\circ}:8^{\circ}]$ (labeled as SFHK-2). This can be understood as the center and the radius of the ring, i.e., $A_r$ and $p_r$, are shifted for different ``born" times from their maximal values. Note that the SFHK-2 yield gets much closer to the TDSE result in which ionization occurs in the whole laser pulse. When ionization is allowed in the SFHK for the whole pulse, we obtain nearly perfect agreement with the TDSE.   

What is the physical origin of the minima in Fig.~\ref{Fig2}(b,d) and in Fig.~\ref{Fig1}(f) in both PMD and the DCS? The QRS simply relates the PMD to the DCS through the factorization. We will show below that these minima are the due to coherent contributions from multiple trajectories with slightly different impact parameters (for DCS in scattering case) or slightly different initial transverse momentum (for PMD in rescattering case) that lead to the same scattering angle. Here, we continue the example shown in Fig.~\ref{Fig2}(a,b). 

First, we show in Fig.~\ref{Fig3}(a) scattering angle $\theta$ vs impact parameter $b$ for laser-free e-Ar$^+$ collisions at incident momentum $k=1.3$~a.u. At this momentum, there are three different trajectories with different impact parameters that leads to the same $\theta$ [except for very small $\theta$ -- see Supplemental Fig.~S3]. In the HK method, the DCS is obtained by adding the contributions from all three trajectories coherently. In particular, the interference between these trajectories results in the minimum near $\theta=140^{\circ}$ for $k=1.3$ a.u. Note that for the Coulomb case, there is only one trajectory for each scattering angle, so no interference can occur. This is why the fully quantum DCS could agree with the classical Rutherford scattering. Note that only $b<3$ a.u. gives contribution to the scattering at $\theta > 90^{\circ}$, as only the electrons that come close to the core can scatter backward. We also show in Fig.~\ref{Fig3}(a) the corresponding case when the laser is on. The two cases are nearly indistinguishable in the figure. For the laser-on case, ``scattering'' angle $\theta$ [defined in Fig.~\ref{Fig2}(b)] vs initial transverse momentum $p_{\perp}$ is shown. Only small initial $p_{\perp}$ contribute to backscattering. The mapping between initial $p_{\perp}$ and the laser-free impact parameter is $b=p_{\perp}\Delta t$, where $\Delta t=0.6T$ can be thought off as the electron traveling time from the born time to the scattering region, with $T$ being the laser period. This result can be interpreted as due to the small effect of the atomic core potential on the electron transverse motion, as $p_{\perp} \ll p_{\parallel}$ for the whole motion before electron scatters from the core, while the atomic core only acts along radial direction. Note that the results on Fig.~\ref{Fig3}(a) were obtained with all the interactions included. An example of three trajectories is shown in Fig.~\ref{Fig3}(b) for the case when $\theta=130^{\circ}$ for both rescattering (with $p_r=1.3$~a.u.) and laser-free (with $k=1.3$~a.u.) cases. They look nearly identical in the scattering region near the core. Note that due to the laser, the final electron direction in the rescattering case is different from it appears in this figure, see Fig.~S4 in the Supplemental Material. 

\begin{figure}
	\centering
        \includegraphics[width=1.0\linewidth]{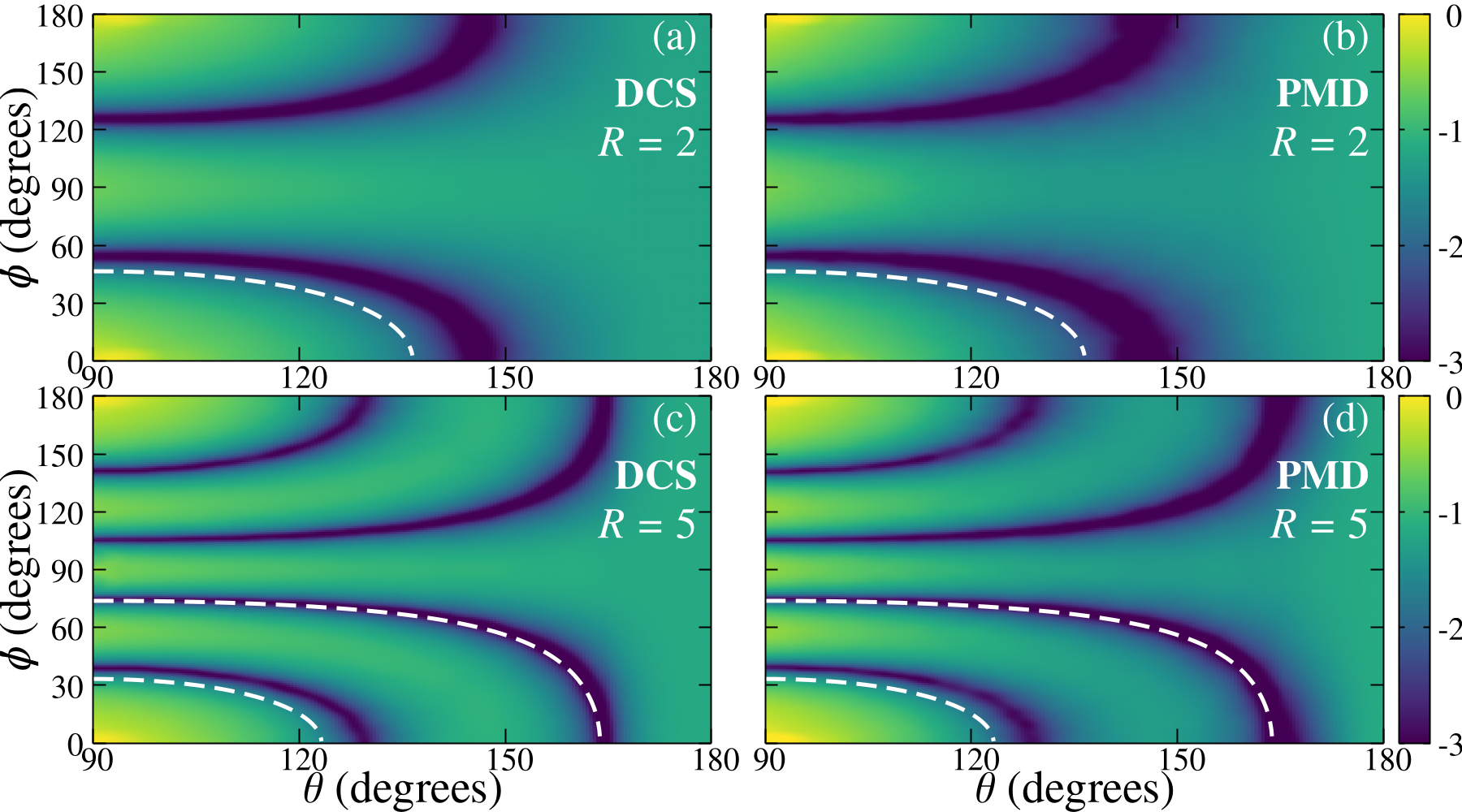}
	\caption{Comparison of (a) the laser-free DCS vs scattering angles of H$_2^+$ with (b) the PMD for electron momentum $p=2.2$~a.u. calculated by the SFHK for internuclear distance $R=2$~a.u.; (c) and (d) the same as (a) and (b), but for $R=5$~a.u. A 800-nm laser pulse with an intensity of $3\times 10^{14}$ W/cm$^2$ was used. The molecule axis is perpendicular to the laser polarization. The white dashed line indicates the position of the two-center interference according to the IAM.} 
	\label{Fig4}
\end{figure}

Putting together, there is one-to-one mapping between rescattering and laser-free scattering trajectories. Each rescattering trajectory carries a weight equal to its ionization probability amplitude which also depends on its initial momentum at the born time. Nevertheless, $p_{\perp}$ distribution relevant for backscattering is very narrow so that the weights are practically the same for all three trajectories -- see Fig.~\ref{Fig3}(a). The total contributions from all trajectories result in a nearly perfect mapping between rescattering yield in the PMD and laser-free scattering DCS that has been employed by the QRS. In this Ar example, there are only three types of trajectories. But the number of trajectories leading to the same final scattering angle could be much larger, see an example for Xe in Fig.~S5 in the Supplemental Material. 

Our method can be extended to molecules in a straightforward manner. For each fixed-in-space molecule, the computational efforts are nearly the same as for atoms, if a single-active-electron model potential is available. Figs.~\ref{Fig4}(a) and (b) compare the PMD with the corresponding DCS for H$_2^+$ vs scattering angles for electron momentum $k=2.2$ a.u. The results, including the positions of the two-center interference minimum, look nearly identical, confirming the validity of the QRS factorization for back-scattered electron with $\theta\gtrapprox 90^{\circ}$. The prediction from the independent atom model (IAM) for the minimum position, shown as the dashed line, is clearly shifted as compared to the SFHK results. The IAM prediction becomes more accurate as the internuclear distance increased to $R=5$ a.u., as shown in Figs.~\ref{Fig4}(c) and (d), as expected. We remark that the accuracy of these interference structures are of critical importance for the LIED ultrafast imaging technique \cite{Blaga:nature12,Wolter:science16,Pullen:NatCom15}.

In conclusion, we propose a method, the SFHK, that is as accurate as the TDSE. Based on this method, we show that there are typically multiple trajectories, which appear in the continuum at the same time, but with slightly different initial transverse momenta, leading to a given final momentum in the high-energy PMD. This is in contrast to the common belief of just two (the long and short) trajectories. Our result is therefore directly relevant for the LIED and other ultrafast imaging techniques that rely on recollision physics. Due to its semiclassical nature, our method can provide intuitive interpretations of various phenomena in strong-field physics, as illustrated here on the example of correspondence between rescattering and laser-free scattering. Although the method was demonstrated only for atoms and H$_2^+$, the extension to polyatomic molecules should be straightforward. We expect that the method can be extended for other recollision phenomena such as high harmonic generation and non-sequential double ionization. Finally, we remark that the SFHK method relies on solving classical Newton equations for independent trajectories. Therefore, primitive parallelization can be easily implemented.

\begin{acknowledgments}
This work was supported by the U.S. Department of Energy (DOE), Office of Science, Basic Energy Sciences (BES) under Award Number DE-SC0023192. Part of the computational work performed on this project was done with help from the Storrs High Performance Computing cluster as well as with services provided by the OSG Consortium \cite{osg06,osg07,osg09}.
\end{acknowledgments}

\appendix*
\setcounter{equation}{0}

\section{Appendix} Consider a target with Hamiltonian $\hat{H}=\hat{H}_0+\hat{V}$, where $\hat{H}_0$ is the laser-free Hamiltonian and $\hat{V}={\bf{r}}\cdot{\bf{E}}(t)$ is the electron-laser interaction in the length gauge. The solution of the TDSE can be written as \cite{Ivanov:jmo05,Spanner:prl03,Lai2015}
\begin{equation}
\begin{aligned}
\Psi \left( {{\bf{r}},{t}} \right) =&  - i\int_{{t_0}}^{{t}} {dt'} \left\langle {{\bf{r}}\left| {{e^{ - i\hat H\left( {t - t'} \right)}}\hat V{e^{ - i{{\hat H}_0}\left( {t' - {t_0}} \right)}}} \right|{\Phi _0}} \right\rangle   \\
 &+ \left\langle {{\bf{r}}\left| {{e^{ - i{{\hat H}_0}(t-t_0)}}} \right|{\Phi _0}} \right\rangle, \qquad \qquad \qquad \qquad \ \ \ \
\end{aligned}
\label{ExactWF}
\end{equation}
where $\left|\Phi _0\right\rangle$ is the initial state. Equation (1) in the main text is obtained by replacing the full propagator $e^{-i\hat{H}(t-t')}$ with the HK propagator. The projection of coherent state to the momentum space and the complex-valued prefactor, can be written, respectively, as ~\cite{Kluk1986,Kay1994_1}
\begin{equation}
\left\langle {{{\bf{p}}_f}}
 \mathrel{\left | {\vphantom {{{\bf{p}}_f} {{{\bf{p}}_t},{{\bf{q}}_t}}}}
 \right. \kern-\nulldelimiterspace}
 {{{{\bf{p}}_t},{{\bf{q}}_t}}} \right\rangle  = {\left( {{ \dfrac{1}{{2\pi \gamma }} }} \right)^{3/4}}\exp \left[ { - { \dfrac{{{{\left( {{\bf{p}}_f - {{\bf{p}}_t}} \right)}^2}}}{{4\gamma }} } - i{\bf{p}}_f{\bf{.}}{{\bf{q}}_t}} \right] ,
\label{GSp}
\end{equation}

\begin{equation}
{C_{{\bf{pq}}t}} = {\left\{ {\det \left[ {{ \dfrac{1}{2} }\left( {{ \dfrac{{\partial {{\bf{p}}_t}}}{{\partial {\bf{p}}}} } + { \dfrac{{\partial {{\bf{q}}_t}}}{{\partial {\bf{q}}}} } - 2i\gamma { \dfrac{{\partial {{\bf{q}}_t}}}{{\partial {\bf{p}}}} } - { \dfrac{1}{{2i\gamma }} }{ \dfrac{{\partial {{\bf{p}}_t}}}{{\partial {\bf{q}}}} }} \right)} \right]} \right\}^{1/2}.}
\label{prefactor}
\end{equation}
The actions are described as \cite{Brennecke2020,Lai2015}
\begin{equation}
S_ \downarrow ^0 = {I_p}{t_s} - \int_{{t_s}}^{{t'}} {d\tau{ \dfrac{{{{\left|{{{\bf{p'}}} + {\bf{A}}\left( \tau \right)} \right|}^2}}}{2} }} ,
\label{Scomplex}
\end{equation}

\begin{equation}
{S_ \to } = \int_{{t'}}^t  {d\tau\left[ {{ \dfrac{{{\bf{p}}_t^2}}{2} } - V_0\left({{{\bf{q}}_t}} \right) - {{\bf{q}}_t}.{\bf{E}}\left( \tau \right)} \right]}
\label{Sreal}
\end{equation}
with ionization energy $I_p$.

Each coherent state wavefunction has a fixed width $\gamma$, in a similar spirit as in the frozen Gaussian method proposed by Heller \cite{Heller:jcp81}. In our calculations, $\gamma$ was chosen to be 0.009. Our results were found to be stable, if this width changes by one to two orders of magnitude. In principle, one might also use multiple widths to accelerate the convergence \cite{Kay1994_1,Kay1994_2}. The filtering technique proposed by Kay \cite{Kay1994_3} was used in our calculations. We have checked that our results are stable with respect to the filtering parameter. Other filtering techniques might be used in the future to further accelerate the convergence. To improve the accuracy of the calculations, we have also projected the final state to the Coulomb waves in Eq.~(2). We found that this procedure leads to only minor improvements for the high-energy PMD. The accuracy can also be improved if the trajectories are propagated further for after the laser is over.

For each atomic target, the single-active electron approximation with model potential $V_0(r)$ was used \cite{Tong:jpb05}. In our calculation, up to about $10^{8}$ trajectories were used. For calculations of the DCS within the SFHK, the initial electron wave-packet is assumed to be a Gaussian centered at $x_0=200$ a.u., incident along the negative $x$-direction with $p_x=1.8$ a.u. toward the ion placed at the origin.

Near $\theta=180^{\circ}$, there is a glory scattering effect associated with the axial caustic singularity, as reported by Xia {\it et al.} \cite{Xia:prl18}. This might lead to a large prefactor $C$ in Eq.~(\ref{SFHKtWF}). To avoid this complication, for $p'_{\perp} < 0.03$ a.u., we make an approximation $C=1$. In other words, we neglect the prefactor (as commonly done in early works by different groups). This approximation leads to some inaccuracies in our results near $\theta=180^{\circ}$, as seen in Fig.~\ref{Fig1}~(e)-(f).


\end{document}